\documentclass[conference]{IEEEtran}
\IEEEoverridecommandlockouts
\usepackage[utf8]{inputenc}
\usepackage[T1]{fontenc}
\usepackage{cite}
\usepackage{amsmath,amssymb,amsfonts}
\usepackage{algorithm}
\usepackage{algpseudocode}
\usepackage{graphicx}
\usepackage{textcomp}
\usepackage{xcolor}
\usepackage{booktabs}
\usepackage{array}
\usepackage{url}
\usepackage{hyperref}
\usepackage{tabularx}
\usepackage{flushend}
\def\BibTeX{{\rm B\kern-.05em{\sc i\kern-.025em b}\kern-.08em
    T\kern-.1667em\lower.7ex\hbox{E}\kern-.125emX}}
\begin{document}

\title{Quantifying the Privacy Posture of Operator-Side 5G/O-RAN Profiles
}

\author{
\IEEEauthorblockN{Nikolaos Kekatos, Apostolos Valiakos, Alexios Lekidis, Elpiniki Papageorgiou}
\IEEEauthorblockA{\textit{University of Thessaly, Department of Energy Systems}\\
Larissa, Greece\\
\{\textit{nkekatos, avaliakos, alekidis, elpinikipapageorgiou}\}\textit{@uth.gr}}
}

\maketitle
{\renewcommand{\thefootnote}{}\footnotetext{\copyright~2026 IEEE. Personal use of this material is permitted. Permission from IEEE must be obtained for all other uses, in any current or future media, including reprinting/republishing this material for advertising or promotional purposes, creating new collective works, for resale or redistribution to servers or lists, or reuse of any copyrighted component of this work in other works. This work has been accepted for publication at the 2026 IEEE International Conference on Cyber Security and Resilience (IEEE CSR 2026).}}

\begin{abstract}
Operator-side network profiles derived from 5G/O-RAN traffic carry personal data such as ephemeral subscriber identifiers, slice-level KPIs, and control-plane signalling, and must be anonymised before release to a federated-learning aggregator, threat-intelligence exchange, or ML training pipeline. We study how much re-identification risk remains after standard operator-side anonymisation. We quantify privacy posture with $k$-anonymity, $\ell$-diversity and $t$-closeness, aggregate them into a composite Privacy-Posture Index (PPI), and measure residual re-identification across eight transformation configurations on internal PCAP captures and the public Idaho Labs 5GAD corpus, under a full-QI syntactic bound and two simulated adversaries. The evaluation is modest in scale, and we read its trends as indicative rather than definitive. Three findings emerge. Pseudonymisation alone leaves re-identification unchanged; material privacy gains arise when quasi-identifiers are coarsened through generalisation, optionally combined with suppression. A downstream classification task then shows that suppression-heavy releases retain majority-class utility but sacrifice much of their minority-class recall, a cost the aggregate metrics hide. Finally, the standard $k_{\min}$-based PPI correlates only modestly with the disclosure bound and not at all with the partial-knowledge attack, whereas a mean-class-size variant $\mathrm{PPI}'$ correlates strongly with all three disclosure/attack measures; we therefore read PPI as a regulator-facing summary, not a security bound. The profiles are produced by passive operator-side monitoring with rule-based DPI; our contribution is the privacy-quantification layer that computes these metrics, applies the transformation policy, and exposes both through an inspectable dashboard.
\end{abstract}


\section{Introduction}
\label{sec:intro}
6G is expected to deepen the data dependency of mobile networks. Cloud-native 5G Cores, disaggregated O-RAN deployments, and ML-driven xApps/rApps continuously ingest control- and user-plane telemetry to optimise slicing, mobility, and security~\cite{porambage2021,nguyen2022}. Much of this telemetry constitutes, or can be linked to, personal data under the GDPR. It may carry subscriber identifiers (IMSI, International Mobile Subscriber Identity; GUTI, Globally Unique Temporary Identifier; SUCI, Subscription Concealed Identifier, a privacy-preserving encoding of the long-term SUPI), session-level KPIs, and behavioural patterns that, alone or with auxiliary information, can enable the singling out or re-identification of subscribers and devices~\cite{khan2020}. Operators therefore face two concurrent obligations. They must \emph{profile} their network, i.e., map 5G core function instances (AMF, Access and Mobility Management Function; SMF, Session Management Function; UPF, User Plane Function), the 5G base station (gNB), and O-RAN control-plane components (Near-RT RIC, Near-Real-Time RAN Intelligent Controller; E2T, E2 Termination)~\cite{polese2023}, detect anomalies, and validate slice isolation. They must also work under a strict data-protection envelope: pseudonymise direct identifiers, bound re-identification risk on quasi-identifiers (QIs), and demonstrate these properties to auditors before any profile leaves the operator~\cite{enisa2019,hintze}.

In practice these obligations have been addressed in isolation. Network-side studies focus on DPI, encrypted-traffic classification, and ML-based intrusion detection for 5G/O-RAN~\cite{aceto2019,polese2023}. Privacy-side studies focus on data-protection models ($k$-anonymity~\cite{sweeney2002}, $\ell$-diversity~\cite{machanavajjhala2007}, $t$-closeness~\cite{li2007}), differential privacy~\cite{dwork2006,dwork2014}, federated learning with DP~\cite{wei2020}, and cryptographic primitives such as homomorphic encryption and zero-knowledge proofs. Tools that do both at once remain rare~\cite{maraj2026,shebaro2011,pang2003}.

We study how much re-identification risk remains in operator-side 5G/O-RAN network profiles after standard syntactic anonymisation, and we characterise it using an exact-QI disclosure bound and two simulated adversaries on internal and external PCAP corpora. The network profiles are produced by PENM (Privacy-Enhanced Network Monitoring), an operator-side monitoring tool built in XTRUST-6G\footnote{XTRUST-6G: Extended Zero-Trust and Intelligent Security for Resilient and Quantum-Safe 6G Networks and Services, an EU Horizon Europe SNS-JU project.}; PENM is standard tooling, not a contribution here; our focus is the privacy-quantification layer that operates on its output. Our contributions are:

\begin{itemize}
\item We quantify the privacy posture of 5G/O-RAN operator-side network profiles using $k$-anonymity, $\ell$-diversity, $t$-closeness, and a composite Privacy-Posture Index (PPI).
\item Residual disclosure is evaluated through an exact-QI syntactic bound and two simulated adversaries, a partial-QI adversary and a probabilistic linkage adversary, on 18 PCAPs (10 internal captures and 8 Normal-1UE captures from the external Idaho Labs 5GAD corpus~\cite{idaholabs5gad}) across eight transformation configurations.
\item The configuration sweep quantifies how QI generalisation and suppression reshape equivalence-class structure and residual disclosure, while a downstream ML experiment measures the resulting task-level utility cost: suppression-heavy releases keep majority-class accuracy but lose about half of their minority-class recall, which an exploratory label-aware policy recovers in the fixed split evaluated here.
\item HMAC pseudonymisation preserves the equivalence-class partition on the QI tuple and therefore yields no measurable privacy gain in this setup; only generalisation of a coarse quasi-identifier (IP-prefix generalisation with threshold suppression, C4, or protocol-count binning, C6) raises mean~$k$ appreciably above~1.
\item We show that PPI's $k_{\min}$ component does not track the partial-knowledge attack, and that a mean-class-size variant $\mathrm{PPI}'$ is strongly associated with all three disclosure/attack measures ($r{=}0.63$--$0.79$, combined corpus).
\end{itemize}

The remainder of the paper is structured as follows. Section~\ref{sec:background} surveys privacy quantification, syntactic models, and pseudonymisation. Section~\ref{sec:profiler} describes the network profiles used as input. Section~\ref{sec:privacy} details the privacy-quantification and transformation layer, our main contribution. Section~\ref{sec:results} reports the experimental results, and Section~\ref{sec:conclusions} concludes.

\section{Background and State of the Art}
\label{sec:background}

\subsection{Privacy quantification: syntactic models}
\label{sub:syntactic}
The empirical motivation for privacy quantification is well established, dating to the foundations of statistical disclosure control~\cite{dalenius1977}. Sweeney showed that $87\%$ of the U.S.\ population is uniquely identifiable from the combination of ZIP code, sex, and date of birth~\cite{sweeney2002}, and de Montjoye \emph{et al.}\ re-identified $95\%$ of individuals in a mobility dataset from only four approximate location--time points~\cite{demontjoye2013}. PENM baselines the three standard syntactic models: $k$-anonymity~\cite{sweeney2002,samarati2001}, which bounds identity-disclosure probability by $1/k$ on chosen quasi-identifiers and admits partition algorithms such as Mondrian~\cite{lefevre2006}; $\ell$-diversity~\cite{machanavajjhala2007}, which blocks homogeneity attacks by requiring at least $\ell$ well-represented sensitive values per class; and $t$-closeness~\cite{li2007}, which bounds the distance between the in-class and the global distribution of the sensitive attribute. These models can be insufficient on their own~\cite{aggarwal2005,kifer2011}, and we therefore treat them as a transparent baseline combined with cryptographic pseudonymisation~\cite{wagner2018,arshad}; differential privacy is left to an offline post-processing layer (Section~\ref{sec:conclusions}).

\subsection{Cryptographic pseudonymisation}
\label{sub:pseudo}
Pseudonymisation is explicitly defined and repeatedly recognised as a data-protection safeguard in the GDPR (Art.~4(5), Recital~26, Art.~25), and ENISA's guidance~\cite{enisa2019} classifies it along three orthogonal axes that are reflected directly in PENM~\cite{limniotis2024,hintze}: \emph{primitive} (keyed hash or HMAC, format-preserving encryption, tokenisation), \emph{determinism} (deterministic vs.\ probabilistic, with the well-known trade-off between linkability and frequency-attack resistance), and \emph{reversibility and trust distribution} (reversible vs.\ irreversible, with controlled re-linkability via additional secret key material). For a network profiler, deterministic HMAC pseudonyms preserve longitudinal analysis on flow records, while an IPv4-shaped pseudonym keeps the data representation compatible with downstream KPI dashboards.

\subsection{Privacy-aware network profiling}
\label{sub:profiling}
Privacy-aware network profiling has historically focused on packet-trace anonymisation (e.g., \texttt{tcpmkpub}~\cite{pang2003}), prefix-preserving IP anonymisation~\cite{shebaro2011}, and encrypted-traffic feature defences~\cite{anderson2017}. Recent work targets feature-layer privacy via MLP-based generalisation~\cite{maraj2026} and FL-based intrusion detection at the edge~\cite{kamatchi2024}. The gap we address is operator-side, inspectable privacy \emph{quantification}: computing the privacy posture of a 5G/O-RAN profile and letting the operator trade utility against re-identification risk before release.

\section{Network Profiles}
\label{sec:profiler}
The privacy layer operates on \emph{network profiles} produced by PENM, an operator-side monitoring tool. From passive packet capture, PENM derives one structured record per host: its inferred 5G/O-RAN role (e.g.\ AMF/gNB, SMF/UPF, Near-RT RIC), obtained by a port- and DPI-based heuristic; the set and count of observed protocols; packet and peer counts; and the UE identifiers (IMSI, GUTI, SUCI) that must be pseudonymised before export. The privacy-quantification layer described next, which is our focus, consumes only these per-host records, not the underlying PCAP.

\section{Privacy-Quantification and Transformation Layer}
\label{sec:privacy}

\subsection{Workflow}
\label{sub:workflow}
The operator workflow follows four steps. (1)~\emph{Ingestion} of a PCAP or a tabular use-case dataset (e.g.\ EV-charging session records). (2)~\emph{Attribute selection}: each field is classified as direct identifier, QI or sensitive attribute (SA), with autogenerated suggestions from the profiling output. (3)~\emph{Transformation policy} (pseudonymisation, generalisation, or suppression). (4)~\emph{Before/after metric reporting} ($k$, $\ell$, $t$) plus textual suggestions.

\subsection{Privacy metrics}
\label{sub:metrics}
For a release $R$ with QI tuple $\mathbf{Q}$ and sensitive attribute $S$, let $\mathcal{E}$ be the partition of $R$ into equivalence classes on $\mathbf{Q}$. PENM computes:
\begin{align}
k(R) &= \min_{E \in \mathcal{E}} |E| \label{eq:k}\\
\ell(R) &= \min_{E \in \mathcal{E}} \big|\{s : s = S(r),\, r \in E,\, s \neq \emptyset\}\big| \label{eq:l}\\
t(R) &= \max_{E \in \mathcal{E}} \tfrac{1}{2}\sum_{v} \big|p_E(v) - p_R(v)\big| \label{eq:t}
\end{align}
where $p_E$ and $p_R$ are the empirical distributions of $S$ on the class $E$ and on $R$. We use total-variation distance (TVD) for $t$, which is bounded in $[0,1]$ and computationally trivial. The module also emits rule-based suggestions: a raw IP QI prompts pseudonymisation and aggregation to a /24 or /16 subnet, low $\ell$ prompts dropping a QI, and $t>0.5$ prompts threshold-based suppression.

\subsection{Transformation primitives}
\label{sub:transforms}
\textbf{Pseudonymisation.} Four techniques are supported. (i)~\emph{HMAC-SHA-256} (default) with a controller-side key derives a hex token (48-bit in the prototype for dashboard compactness; production should use 96--128 bits to keep collision probability negligible at scale); this prevents practical offline dictionary testing by a recipient without the controller-held key, provided the key is generated, stored, and rotated securely. (ii)~\emph{Salted SHA-256} is a non-keyed, integrity-oriented alternative; it should not be used for identifiers from small or enumerable domains, for which keyed HMAC is preferred. (iii)~\emph{Random tokenisation} replaces values with fresh random tokens whose mapping is held in memory only for the request, breaking offline linkability. (iv)~\emph{IPv4-shaped HMAC pseudonymisation} keeps the first octet bucket and replaces the remaining three octets with the leading bytes of an HMAC over the original address, producing a valid IPv4 string for downstream tools. It is a format-compatible heuristic, not an AES-FF1 or prefix-preserving construction. Three scope policies are available: \emph{deterministic} (cross-run linkable), \emph{per-pilot} (linkable within a pilot, isolated across pilots), and \emph{per-run} (fresh per request).

\textbf{Generalisation.} Configurable profiles are applied per column: IP-prefix generalisation to /24 or /16; fixed-width numeric binning (e.g., bin widths of 5 or 10); text-prefix masking that retains only the first one or three characters; and protocol-family abstraction that collapses transport/application protocols into SCTP, TCP, UDP, and service-discovery families.

\textbf{Suppression.} Direct suppression replaces the value with the symbol \texttt{*}. Threshold-based suppression takes a per-column integer $\tau$ and replaces values whose support across the dataset is below $\tau$; this prevents rare values from pinning $k$ at 1.

\subsection{Privacy-Posture Index (PPI)}
\label{sub:ppi}
The triple $(k,\ell,t)$ is informative but difficult to communicate to a
non-technical regulator and hard to compare across pipelines. We
therefore define a composite \emph{Privacy-Posture Index} that
collapses the three metrics into a single $[0,1]$ score, in the spirit
of how the Common Vulnerability Scoring System (CVSS) aggregates vulnerability dimensions~\cite{cvss40}:
Let $s_k = \min(k / k^\star, 1)$, $s_\ell = \min(\ell / \ell^\star, 1)$, and $s_t = 1 - t$, with target thresholds $k^\star$ and $\ell^\star$. The PPI is then defined as the convex combination
\begin{equation}
\mathrm{PPI}(k,\ell,t) \;=\; w_k \, s_k + w_\ell \, s_\ell + w_t \, s_t,
\label{eq:ppi}
\end{equation}
with $w_k, w_\ell, w_t \in [0,1]$ and $w_k + w_\ell + w_t = 1$, so that $\mathrm{PPI} \in [0,1]$. We use $k^\star{=}5$ and $\ell^\star{=}2$ as illustrative operational targets commonly cited in practice~\cite{enisa2019}; they are \emph{not} regulatory thresholds and do not by themselves establish anonymisation or GDPR compliance. The default weights $(w_k, w_\ell, w_t){=}(0.4, 0.3, 0.3)$ place more weight on $k$ because identity disclosure is the primary regulator-facing risk; they are conventional, not tuned.

The $s_k$ component uses $k_{\min}$, the smallest class size, which captures the single worst-case record but ignores the rest of the class-size distribution, the part an \emph{averaged} attack sees. We therefore also define $\mathrm{PPI}'$, identical to Eq.~\ref{eq:ppi} except that $s_k'=\min(k_{\mathrm{rw}}/k^\star,1)$ uses the record-weighted mean class size $k_{\mathrm{rw}}=\tfrac{1}{N}\sum_{E}|E|^2$ (the class size a typical record hides in), distinct from the minimum $k_{\min}$. Section~\ref{sec:results} reports both and measures how each correlates with residual re-identification.

\subsection{Operator interface}
\label{sub:api}
\begin{figure}[t]
\centering
\includegraphics[width=0.72\linewidth]{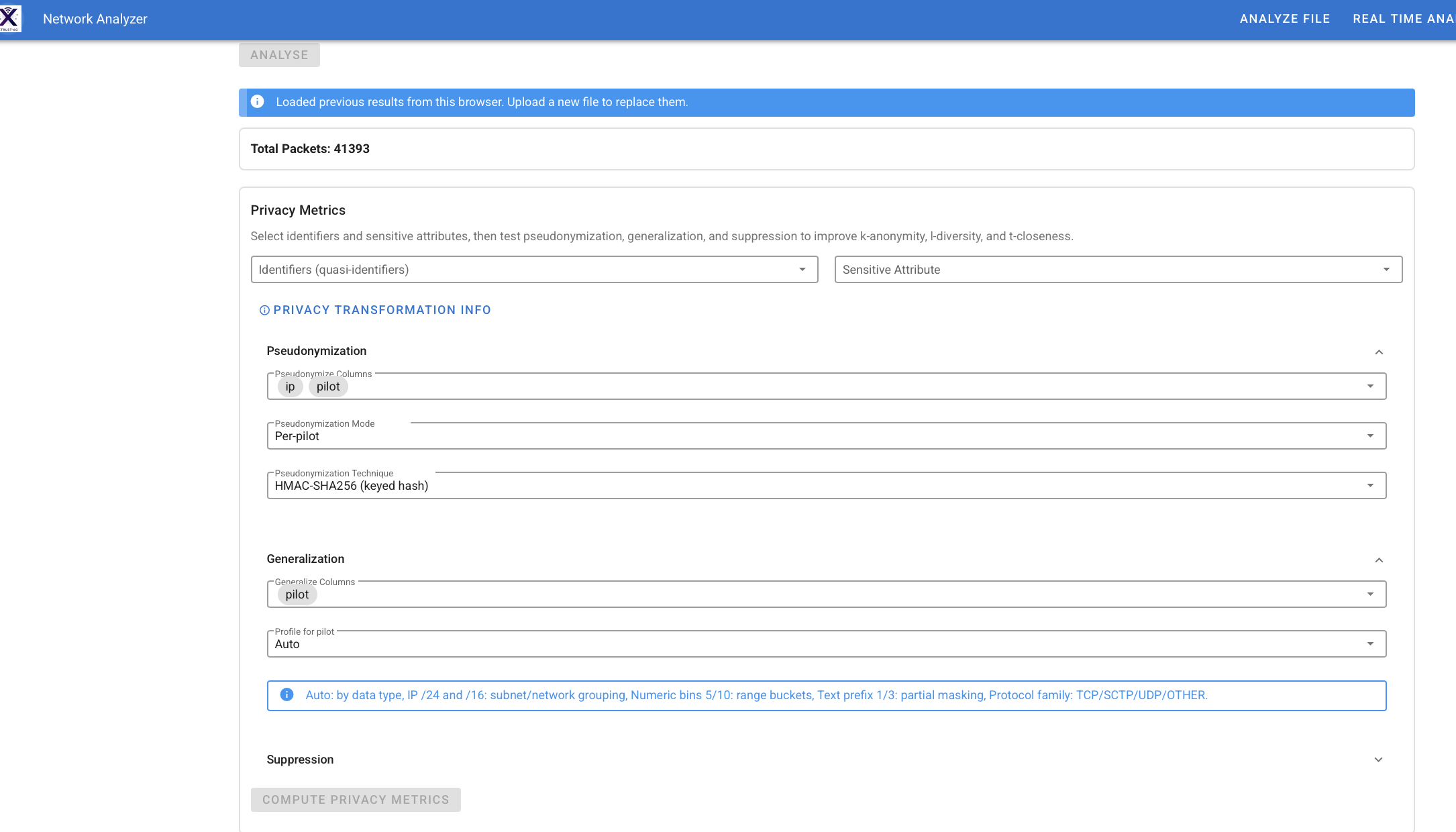}\\[2pt]
\includegraphics[width=0.72\linewidth]{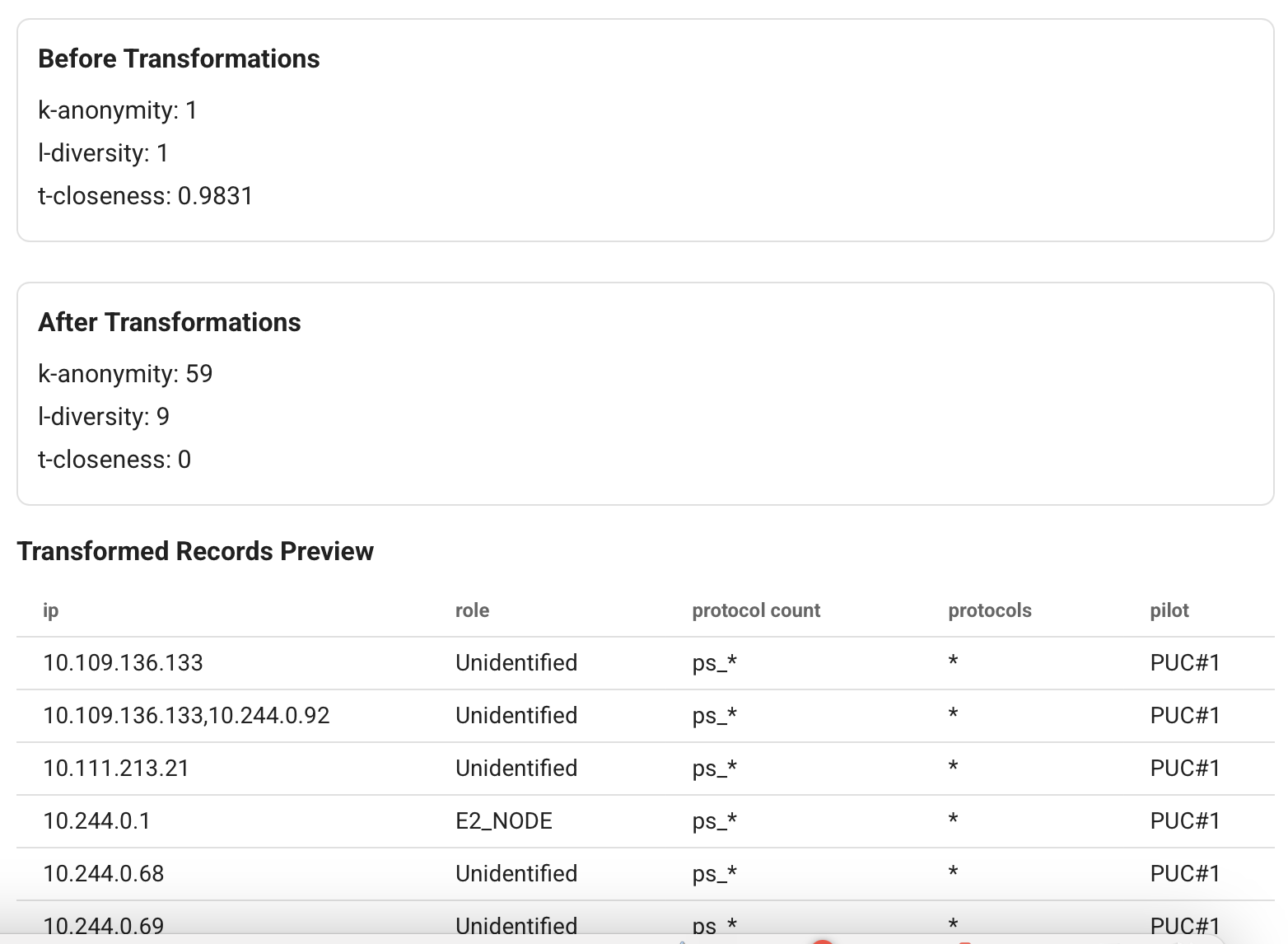}
\caption{The PENM privacy dashboard. \emph{Top:} per-column transformation controls. \emph{Bottom:} before/after $(k,\ell,t)$ and transformed-records preview. The values shown ($k{:}\,1{\to}59$, $t{:}\,0.98{\to}0$) are an illustrative dashboard demonstration on an EV-charger PCAP and are not part of the experimental evaluation (see the reproducible result in Section~\ref{sub:puc1}).}
\label{fig:dashboard}
\end{figure}

The privacy layer is exposed through a web dashboard: the operator selects identifiers, quasi-identifiers and the sensitive attribute, configures the transformation policy, and inspects the before/after $(k,\ell,t)$, the rule-based suggestions, and a preview of the transformed records side by side. Each transformation and its effect on the metrics is visible rather than hidden behind a single anonymisation action. Figure~\ref{fig:dashboard} shows it in operation: the upper panel exposes the per-column policy, the lower panel the before/after metrics and the exportable transformed-records preview.

\section{Experimental Evaluation}
\label{sec:results}

\subsection{Threat model}
\label{sub:threat}
We assume a trusted data controller (the operator running PENM) and an untrusted recipient that obtains a released profile after the operator-side privacy transformation. Recipients of interest include downstream analytics pipelines such as a federated-learning aggregator, a cyber-threat-intelligence exchange, or an offline ML training stage. The adversary's goal is record-level re-identification: linking a released row to an individual subscriber, device, or function instance. We consider one analytical disclosure bound and two simulated adversaries over the chosen quasi-identifiers. The \emph{exact-QI} setting assumes knowledge of every QI value of every target record and gives the full-QI syntactic disclosure bound $1/|E|$. The \emph{partial-QI} adversary knows each QI value with probability $1{-}p$ (we use $p{=}0.5$), modelling auxiliary-data leaks. A stronger probabilistic \emph{linkage} adversary is defined in Section~\ref{sub:adversary}. Across all evaluated settings, the recipient is assumed to know the transformation policy $\Pi$ and the QI/SA selection (Kerckhoffs-style); only the controller-held HMAC key and the original records are unknown. A successful re-identification is one where the adversary's residual candidate set has size 1 and matches the true row; for larger candidate sets we credit the adversary a probability $1/|E|$ of a correct guess.

\subsection{Setup and datasets}
\label{sub:setup}
We evaluate PENM on three datasets. \textbf{(i)~O-RAN testbed}: 12 sessions captured by the passive sensor and tagged by the rule-based role classifier (E2 Node, Near-RT RIC, E2T). Quasi-identifiers are the RIC Region (R1/R2), a 5-minute time window, and the inferred node role. The sensitive attribute is the slice differentiator (\texttt{SD}) bucket within S-NSSAI: $G1{=}$~eMBB cloud-gaming, $T1{=}$~URLLC telemedicine, $M1{=}$~mMTC smart-metering. \textbf{(ii)~EV-charging dataset}: 513 sessions with fields \emph{Connector Chargepoint Name}, \emph{Start\_Time}, \emph{Last\_Time}, \emph{Duration}, \emph{Total kWh}, \emph{Id\_Tag}, \emph{Start\_Wh}, \emph{Last\_Wh}, \emph{Transaction Status/ID}, \emph{Connector Standard/ID/Power Type}. We pick QI = \{\emph{Station}, \emph{Start\_Time}, \emph{Connector ID}\} and SA = \emph{Total kWh} (capped at the 99th percentile, $29.254$ kWh, for robust $t$). \textbf{(iii)~Internal PCAP corpus}: ten PCAPs (an operator-network capture, an EV-charging OCPP capture, an O-RAN testbed capture, and seven smaller testbed captures of 2--5 IPs each). The per-PCAP record extraction uses tshark with payload stripping; the resulting per-IP record schema is \{ip, role, protocols, protocol\_count, packets, peer\_count, pilot\}. Quasi-identifiers and sensitive attributes vary by configuration (C0--C7); the full setup is summarised in Table~\ref{tab:setup}.

\begin{table*}[t]
\centering
\caption{Reproducibility summary of experimental setups.}
\label{tab:setup}
\renewcommand{\arraystretch}{1.0}
\setlength{\tabcolsep}{5pt}
\footnotesize
\begin{tabularx}{\linewidth}{@{}l >{\hsize=1.45\hsize}X >{\hsize=0.9\hsize}X >{\hsize=0.75\hsize}X >{\hsize=0.9\hsize}X@{}}
\toprule
\textbf{Setup} & \textbf{Records} & \textbf{QI} & \textbf{SA} & \textbf{Transform}\\
\midrule
O-RAN testbed     & 12 sessions, 3 equiv.~classes                                 & RIC region, 5-min window, role     & S-NSSAI SD bucket               & none (intrinsic baseline)\\
EV-charging       & 513 sessions $\to$ 70 retained                                & Station, Start\_Time, Connector ID & Total kWh (10-bucket quantile)  & HMAC on Id\_Tag, hour-bucket Start\_Time, $\tau{=}5$ suppression\\
Internal PCAP corpus   & 10 PCAPs, 2--85 IPs each                                      & varies (C0--C7)                    & varies                          & C0--C7 (Tab.~\ref{tab:sweep})\\
External 5GAD corpus   & 8 5GAD Normal-1UE captures (3 full, 5 200\,k-packet slices), 5--322 IPs each & varies (C0--C7)                    & varies                          & C0--C7 (Tab.~\ref{tab:external})\\
\bottomrule
\end{tabularx}
\end{table*}

\subsection{O-RAN testbed results}
\label{sub:oran}
On the 12-session O-RAN scenario the equivalence-class partition on QI = (Region, Time-window, Role) yields three classes:
\begin{itemize}
\item $C_1 = ($R1$, \, 10\!\!:\!\!00$--$10\!\!:\!\!04, \, $E2 Node$)$, $|C_1|{=}6$, SD\,$\in\{G1,T1,M1\}$.
\item $C_2 = ($R2$, \, 10\!\!:\!\!00$--$10\!\!:\!\!04, \, $E2 Node$)$, $|C_2|{=}3$, SD\,$\in\{G1,T1,M1\}$.
\item $C_3 = ($R1$, \, 10\!\!:\!\!05$--$10\!\!:\!\!09, \, $Near-RT RIC$)$, $|C_3|{=}3$, SD\,$\in\{G1,M1\}$.
\end{itemize}
Hence $k = \min(6,3,3) = 3$, $\ell = \min(3,3,2) = 2$ (the dataset is 2-diverse but not 3-diverse because $C_3$ only sees $\{G1,M1\}$), and $t = \max_{i} \mathrm{TVD}(p_{C_i}, p_R) = 0.25$. Under the exact-QI disclosure model, this release is more defensible than the raw release: identity disclosure is bounded by $1/3$, the attribute-disclosure homogeneity attack is partially mitigated, and the per-class slice-mix deviates from the global distribution by at most $25\%$ TVD. We make no formal regulatory claim from these values alone. We stress that with only 12 sessions in three classes of sizes~6,~3 and~3 this is an illustrative worked example, not a sample from which statistically significant conclusions can be drawn; every quantitative trend claimed in this paper rests on the 144-release configuration sweep of Sections~\ref{sub:adversary}--\ref{sub:external}, not on this snapshot. Raising $k$ (e.g.\ by widening the time window) or $\ell$ (e.g.\ by grouping URLLC and mMTC slice-types) further would not require any change to the profiling pipeline; only the transformation policy.

\subsection{EV-charging results}
\label{sub:puc1}
On the 513-session EV-charging dataset, the original release has $(\text{rows}, k, \ell, t) = (513, 1, 1, 0.90)$; after the transformation policy described below this becomes $(70, 5, 2, 0.73)$, where $t$ is computed on the same 10-bucket quantile binning of the SA in both cases. These values are reproducible from the structured export with our released script. The transformation policy applied was: (1)~\emph{pseudonymisation}: \emph{Id\_Tag} replaced by a deterministic HMAC-SHA-256 token, preserving longitudinal analysis within the dataset scope; (2)~\emph{generalisation}: \emph{Start\_Time} bucketed to the hour (\texttt{numeric\_bins\_10} on the hour-of-day component), \emph{Connector ID} retained at coarse granularity, and \emph{Total kWh} bucketed into quantile bins for stable $\ell$/$t$; and (3)~\emph{suppression}: equivalence classes of size $<5$ suppressed via a threshold $\tau{=}5$ on the equivalence-class support.

The baseline release is highly re-identifying ($k{=}1$ with 513 singleton classes and $t{=}0.90$ on a $10$-bucket quantile binning of \emph{kWh}, applied to make $t$ directly comparable to Eq.~\ref{eq:t}). The transformation raises $k$ to~5 (identity disclosure bounded by $1/5$), raises $\ell$ from~1 to~2, and reduces $t$ from~$0.90$ to~$0.73$. The utility cost is a reduction from $513$ to $70$ rows ($\sim$14\% retained). Whether a release this sparse still supports downstream machine learning is exactly the concern we quantify in Section~\ref{sub:mlutility}. The release is more defensible than the raw release under the stated adversary model and aligned with the ENISA pseudonymisation criteria~\cite{enisa2019}; we make no regulatory-threshold claim without an external assessment.

\subsection{Adversary-success simulation and PPI validation}
\label{sub:adversary}
Reporting $(k,\ell,t)$ alone characterises what the release \emph{claims}; it does not measure what an attacker actually achieves. To complement this in light of Aggarwal's curse-of-dimensionality result~\cite{aggarwal2005} and Kifer--Machanavajjhala's no-free-lunch theorem~\cite{kifer2011}, we evaluate one analytical disclosure bound and two simulated adversaries across a 10-PCAP $\times$ 8-config sweep ($N{=}80$ releases). The \emph{exact-QI} bound assumes knowledge of every QI value; the expected re-identification per row is $1/|E|$, where $E$ is the equivalence class on the QI tuple. This is the full-QI syntactic disclosure bound, a function of the equivalence-class structure that $k$ also measures, not an independent attack. The \emph{partial-QI attack} forgets each QI value with probability $p{=}0.5$, matches the residual query against the release, and guesses uniformly among candidates. The \emph{linkage attack} is stronger: it observes each released QI attribute correctly with probability $0.8$ (else a random value from that attribute's release domain), scores every candidate by the number of matching attributes, and picks uniformly among the top-scoring set, i.e.\ a probabilistic record-linkage adversary rather than exact-subset matching. Both simulated attacks are averaged over 20 fixed-seed Monte-Carlo repetitions with every released row as a target. For the partial-QI attack, a query in which all QI values are forgotten falls back to a uniform $1/N$ guess, and a non-empty residual query that matches no released record receives zero credit; the linkage attack always selects uniformly among the highest-scoring candidates.

The 10 PCAPs span the three scenarios used in this paper (an operator-network capture, an EV-charging OCPP capture, and an O-RAN testbed capture) plus seven small captures from the same testbed environment (2--5 IPs each). Table~\ref{tab:sweep} aggregates results per configuration (average across all 10 PCAPs).

\begin{table*}[t]
\centering
\caption{Internal PCAP corpus: 10 PCAPs $\times$ 8 configurations ($N{=}80$). Cells are mean\,$\pm$\,population sd across PCAPs. $\bar{U}{=}|\mathcal{E}|/N$ is the equivalence-class granularity ratio, which equals the expected exact-QI disclosure probability averaged over records (hence exactReID$\,{=}\,100\,\bar{U}$) and is not an independent utility metric; $\bar{k}$ is the mean per-release minimum class size ($k_{\min}$). exactReID is the full-QI syntactic disclosure bound and partReID the simulated partial-QI attack (\%) (Section~\ref{sub:threat}). Linkage-attack results are omitted for space; their correlations with PPI and $\mathrm{PPI}'$ appear in Table~\ref{tab:ppicorr}.}
\label{tab:sweep}
\renewcommand{\arraystretch}{1.0}
\setlength{\tabcolsep}{6pt}
\footnotesize
\begin{tabular}{@{}lccccc@{}}
\toprule
\textbf{Config} & $\bar{k}$ & $\overline{\mathrm{PPI}}$ & $\bar{U}$ & $\overline{\text{exactReID}}$ (\%) & $\overline{\text{partReID}}$ (\%)\\
\midrule
C0 None                 & $1.00\,{\pm}\,0.00$ & $0.31\,{\pm}\,0.08$ & $1.00\,{\pm}\,0.00$ & $100.0\,{\pm}\,0.0$ & $57.5\,{\pm}\,6.3$\\
C1 HMAC-IP              & $1.00\,{\pm}\,0.00$ & $0.31\,{\pm}\,0.08$ & $1.00\,{\pm}\,0.00$ & $100.0\,{\pm}\,0.0$ & $57.5\,{\pm}\,6.3$\\
C2 IP/24                & $1.10\,{\pm}\,0.30$ & $0.33\,{\pm}\,0.10$ & $0.60\,{\pm}\,0.08$ & $59.5\,{\pm}\,7.6$  & $38.9\,{\pm}\,8.1$\\
C3 IP/16                & $1.10\,{\pm}\,0.30$ & $0.33\,{\pm}\,0.10$ & $0.58\,{\pm}\,0.08$ & $57.7\,{\pm}\,7.9$  & $38.1\,{\pm}\,9.0$\\
C4 IP/24+supp           & $\mathbf{3.00\,{\pm}\,1.26}$ & $\mathbf{0.66\,{\pm}\,0.20}$ & $0.28\,{\pm}\,0.16$ & $28.0\,{\pm}\,16.3$ & $24.0\,{\pm}\,14.6$\\
C5 \{protos,cnt\}       & $1.10\,{\pm}\,0.30$ & $0.32\,{\pm}\,0.10$ & $0.75\,{\pm}\,0.29$ & $74.8\,{\pm}\,29.4$ & $53.9\,{\pm}\,23.0$\\
C6 protoCnt/bin5        & $2.50\,{\pm}\,1.57$ & $0.61\,{\pm}\,0.28$ & $0.25\,{\pm}\,0.17$ & $\mathbf{25.3\,{\pm}\,16.8}$ & $\mathbf{22.3\,{\pm}\,14.8}$\\
C7 role\,$\to$\,protos  & $1.50\,{\pm}\,0.50$ & $0.38\,{\pm}\,0.10$ & $0.42\,{\pm}\,0.20$ & $41.6\,{\pm}\,20.1$ & $30.1\,{\pm}\,15.8$\\
\bottomrule
\end{tabular}
\end{table*}

Four observations follow from Table~\ref{tab:sweep}.

\emph{(i) Pseudonymisation alone does not reduce re-identification.} C1 (HMAC pseudonyms on \texttt{ip}) matches C0 on every privacy metric and disclosure/attack measure. The pseudonyms remain unique, so the equivalence-class structure on the QI tuple is unchanged. This result supports the ENISA position that pseudonymisation must be combined with other PETs~\cite{enisa2019}, and is at odds with the assumption that hashing identifiers is sufficient for anonymisation.

\emph{(ii) Generalisation alone is also insufficient.} C2 and C3 (IP-prefix generalisation to /24 and /16, respectively) reduce exact-QI re-identification to $\sim$60\%, but cannot raise the mean per-release $k_{\min}$ above 1.1 because long-tail singletons persist. Adding threshold suppression to the /24 generalisation (C4 vs.\ C2) gives the largest reduction in this ablation pair: it raises the mean per-release $k_{\min}$ to 3 and cuts exact-QI re-identification by $\sim$72 percentage points relative to baseline.

\emph{(iii) The strongest configuration is QI- and SA-aware rather than transformation-intensive.} C6 (binning a coarse QI such as protocol count) yields the lowest residual re-identification (25.3\% exact, 22.3\% partial) with only modest transformation cost.

\emph{(iv) The exact-QI bound and equivalence-class granularity coincide.} The proxy $\bar{U}{=}|\mathcal{E}|/N$ in Table~\ref{tab:sweep} equals the expected exact-QI re-identification probability, since $\tfrac{1}{N}\sum_{r}1/|E(r)|=|\mathcal{E}|/N$. We therefore treat $\bar{U}$ as a granularity ratio, not an independent utility metric; the genuine, task-level utility cost is measured separately by the downstream classifier of Section~\ref{sub:mlutility}.

\begin{figure}[t]
\centering
\includegraphics[width=\linewidth]{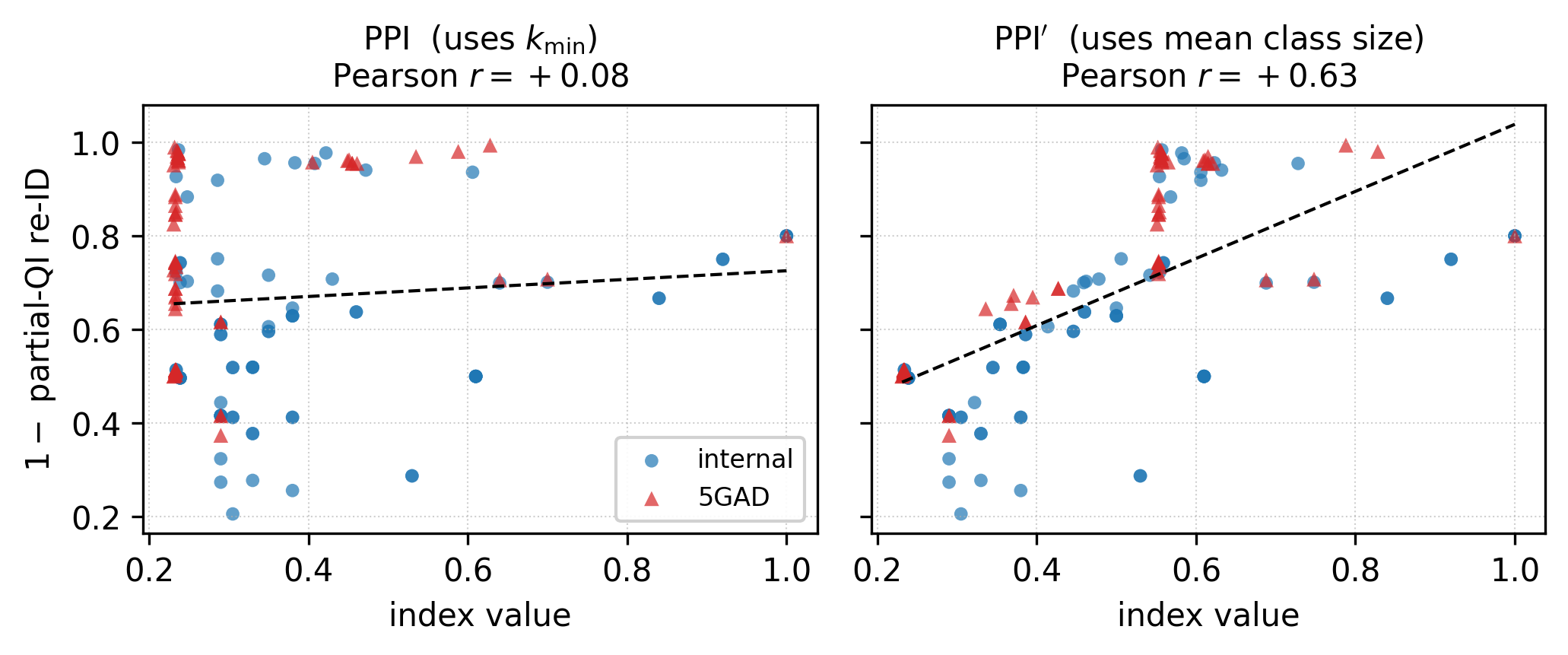}
\caption{Each point is one release ($N{=}144$; internal and 5GAD). PPI ($k_{\min}$, left) is essentially uncorrelated with resistance to the partial-QI attack ($r{=}{+}0.08$); $\mathrm{PPI}'$ ($k_{\mathrm{rw}}$, right) correlates with it ($r{=}{+}0.63$). Dashed line: least-squares fit.}
\label{fig:ppival}
\end{figure}

\emph{PPI validity.} Table~\ref{tab:ppicorr} correlates PPI and $\mathrm{PPI}'$ with $1{-}\text{reID}$ for each disclosure/attack measure, with bootstrap 95\% confidence intervals ($B{=}2000$ resamples). The $k_{\min}$-based PPI correlates only modestly with the exact-QI disclosure bound (combined $r{=}{+}0.30$) and is essentially uncorrelated with the partial-QI attack ($r{=}{+}0.08$, $p{=}0.33$; Fig.~\ref{fig:ppival}, left). The mismatch is structural: PPI rewards $k_{\min}$, whereas the simulated attacks depend on the broader class-size and candidate-set distributions and the exact-QI average equals $|\mathcal{E}|/N$. The variant $\mathrm{PPI}'$ is far more strongly associated with all three measures (combined $r{=}{+}0.79$ exact, ${+}0.63$ partial, ${+}0.75$ linkage; Fig.~\ref{fig:ppival}, right), with confidence intervals disjoint from those of PPI. We keep PPI as the dashboard's conservative $k_{\min}$-oriented indicator and report $\mathrm{PPI}'$ as an attack-alignment variant, not a replacement regulatory metric; because $k_{\mathrm{rw}}$ is dominated by large classes, $\mathrm{PPI}'$ can mask singleton tails and should be read alongside $k_{\min}$. Neither replaces the adversary simulations, which we report alongside.

\begin{table}[t]
\centering
\caption{Pearson correlation of PPI and $\mathrm{PPI}'$ with $1{-}\text{reID}$ (combined corpus, $N{=}144$; bootstrap 95\% CI). $\mathrm{PPI}'$ ($k_{\mathrm{rw}}$) correlates strongly with resistance to all three disclosure/attack measures; PPI ($k_{\min}$) does not.}
\label{tab:ppicorr}
\renewcommand{\arraystretch}{1.1}
\setlength{\tabcolsep}{5pt}
\footnotesize
\begin{tabular}{@{}lcc@{}}
\toprule
Disclosure/attack measure & PPI ($k_{\min}$) & $\mathrm{PPI}'$ ($k_{\mathrm{rw}}$)\\
\midrule
exact-QI bound & $0.30$ $[0.19,0.41]$ & $\mathbf{0.79}$ $[0.74,0.84]$\\
partial-QI     & $0.08$ $[\!-\!0.04,0.19]$ & $\mathbf{0.63}$ $[0.56,0.71]$\\
linkage        & $0.24$ $[0.12,0.35]$ & $\mathbf{0.75}$ $[0.70,0.81]$\\
\bottomrule
\end{tabular}
\end{table}

\emph{Transformation cost.} On an Apple M-series machine (single thread, 5-run medians on synthetic records matching the profile schema), the privacy transformation under configuration C4 (generalisation plus threshold suppression) scales linearly with record count: $10^2$, $10^3$, $10^4$ records process in $0.3$, $3.2$, and $32$\,ms respectively, i.e.\ ${\approx}\,3.2$\,$\mu$s per record or ${\approx}\,3{\times}10^5$ records/s. Pseudonymisation-only (C1) is ${\approx}\,1.5{\times}$ faster. At our record volumes the transformation stage is therefore well below the per-PCAP DPI/profiling cost. This covers the privacy-quantification stage, which is our focus; the profiler's own footprint is standard tooling and out of scope.

\subsection{External validation on 5GAD}
\label{sub:external}
The datasets used so far were produced in-house (testbed captures and an EV-charging record set). To assess whether the results generalise beyond the local environment, we replicate the sweep on 5GAD~\cite{idaholabs5gad}, an independent 5G dataset published by Idaho National Laboratory and based on Open5GS and UERANSIM captures of 1-UE benign traffic. We use eight Normal-1UE PCAP profiles (200,000-packet slices for the larger captures to bound parsing runtime, full file for the shorter ones), each yielding 5--322 records. The captures are comparable in size to the internal scenarios but are produced by a separate testbed, with no shared topology, addressing plan, or role-classifier ground truth.

Table~\ref{tab:external} compares internal (10 PCAPs) and external (8 PCAPs) averages on the same 8 configurations.

\begin{table*}[t]
\centering
\caption{Internal vs external (5GAD) means\,$\pm$\,sd per configuration. exactReID (\%) is the full-QI syntactic disclosure bound, computed independently for each corpus. Lowest per column in bold.}
\label{tab:external}
\renewcommand{\arraystretch}{1.0}
\setlength{\tabcolsep}{6pt}
\footnotesize
\begin{tabular}{@{}lcccccc@{}}
\toprule
& \multicolumn{3}{c}{\textbf{Internal} (10 PCAPs)} & \multicolumn{3}{c}{\textbf{External 5GAD} (8 PCAPs)}\\
\cmidrule(lr){2-4}\cmidrule(lr){5-7}
\textbf{Config} & $\bar{k}$ & PPI & exactReID (\%) & $\bar{k}$ & PPI & exactReID (\%)\\
\midrule
C0 None                 & $1.00\,{\pm}\,0.0$ & $0.31\,{\pm}\,0.08$ & $100.0\,{\pm}\,0.0$ & $1.00\,{\pm}\,0.0$ & $0.24\,{\pm}\,0.02$ & $100.0\,{\pm}\,0.0$\\
C1 HMAC-IP              & $1.00\,{\pm}\,0.0$ & $0.31\,{\pm}\,0.08$ & $100.0\,{\pm}\,0.0$ & $1.00\,{\pm}\,0.0$ & $0.24\,{\pm}\,0.02$ & $100.0\,{\pm}\,0.0$\\
C2 IP/24                & $1.10\,{\pm}\,0.3$ & $0.33\,{\pm}\,0.10$ & $59.5\,{\pm}\,7.6$  & $1.00\,{\pm}\,0.0$ & $0.24\,{\pm}\,0.02$ & $64.9\,{\pm}\,5.1$\\
C3 IP/16                & $1.10\,{\pm}\,0.3$ & $0.33\,{\pm}\,0.10$ & $57.7\,{\pm}\,7.9$  & $1.00\,{\pm}\,0.0$ & $0.24\,{\pm}\,0.02$ & $51.8\,{\pm}\,6.9$\\
C4 IP/24+supp           & $3.00\,{\pm}\,1.3$ & $0.66\,{\pm}\,0.20$ & $28.0\,{\pm}\,16.3$ & $3.00\,{\pm}\,0.5$ & $0.48\,{\pm}\,0.07$ & $11.4\,{\pm}\,10.8$\\
C5 \{protos,cnt\}       & $1.10\,{\pm}\,0.3$ & $0.32\,{\pm}\,0.10$ & $74.8\,{\pm}\,29.4$ & $1.00\,{\pm}\,0.0$ & $0.24\,{\pm}\,0.02$ & $30.5\,{\pm}\,26.8$\\
C6 protoCnt/bin5        & $2.50\,{\pm}\,1.6$ & $0.61\,{\pm}\,0.28$ & $\mathbf{25.3\,{\pm}\,16.8}$ & $1.88\,{\pm}\,1.4$ & $0.42\,{\pm}\,0.27$ & $\mathbf{5.0\,{\pm}\,5.7}$\\
C7 role\,$\to$\,protos  & $1.50\,{\pm}\,0.5$ & $0.38\,{\pm}\,0.10$ & $41.6\,{\pm}\,20.1$ & $1.12\,{\pm}\,0.3$ & $0.29\,{\pm}\,0.15$ & $10.0\,{\pm}\,11.4$\\
\bottomrule
\end{tabular}
\end{table*}

Three findings transfer from the internal corpus to 5GAD.

\emph{Result 1: the configuration ranking is preserved.} C4 and C6 outperform the other configurations on both corpora; C0 (no transformation) and C1 (pseudonymisation alone) coincide on every metric, as HMAC pseudonyms preserve the equivalence-class partition by construction (Section~\ref{sub:limits}).

\emph{Result 2: 5GAD profiles are more amenable to anonymisation than the internal profiles.} On 5GAD, C6 reduces exact-QI re-identification to $5.0\,\%$ and C4 to $11.4\,\%$. The internal corpus gives $25.3\,\%$ and $28.0\,\%$ for the same configurations. We attribute the gap to dataset structure. 5GAD captures aggregate $\sim$200\,k packets across $\sim$100 IPs, so the per-IP protocol-count and protocol-set distributions are richer and the QI/SA partitions are larger. Larger dominant classes increase the record-weighted mean class size $k_{\mathrm{rw}}$ and reduce expected exact-QI disclosure, even where $k_{\min}$ remains low because of singleton tails. This is consistent with Aggarwal's analysis~\cite{aggarwal2005}: when each QI bucket contains enough rows, syntactic anonymisation is more effective. It also supports the interpretation that the small-PCAP $t{=}0$ outcomes visible in Table~\ref{tab:sweep} are a small-sample artefact, not a property of the transformation pipeline.

\emph{Result 3: 5GAD has mean $k_{\min}$ no larger than the internal corpus, yet lower exact-QI re-identification.} On its face this looks contradictory. The resolution is that $k_{\min}$ (the smallest class size) and the record-weighted mean $k_{\mathrm{rw}}$ measure different things. Expected exact-QI success equals $|\mathcal{E}|/N$ and is empirically associated with $k_{\mathrm{rw}}$, the class size a typical record hides in, rather than with $k_{\min}$. On 5GAD, $k_\text{min}$ stays at $1$ because a few management IPs form singleton classes, but most records lie in much larger classes than in the internal captures. This is a known limitation of $k$-anonymity~\cite{aggarwal2005,kifer2011}, and is why we report PPI and residual re-identification side by side: $k_{\min}$ alone misranks releases when the tail behaves differently.

Across the combined 144-release corpus PPI correlates with $1{-}\text{exactReID}$ at $r{=}{+}0.30$, below the internal-only $r{=}{+}0.45$, because its $k$-component rewards $k_{\min}$, whereas the exact-QI average equals $|\mathcal{E}|/N$ and the simulated attacks depend on the broader class-size distribution. The $\mathrm{PPI}'$ variant weighted by mean class size (Section~\ref{sub:adversary}) closes this gap.

\subsection{Downstream ML utility under suppression}
\label{sub:mlutility}
A release that retains only $\sim$14\,\% of its rows, as in the EV case, raises a fair question: does what survives still support a downstream ML task, and if so, at what cost to the rare, ``edge'' cases? The structured EV set is too small and singular to answer this on its own, so we run the test on the pooled PCAP corpus, which lets us drive the surviving fraction across its full range. We pool 664 per-IP profile records (the internal corpus plus the 5GAD 200\,k-packet slices) and define a downstream task: predict node \emph{role} from the behavioural features \{\texttt{protocol\_count}, \texttt{packets}, \texttt{peer\_count}\} (excluding the protocol set, which fixes the role by construction), collapsing roles rarer than six records into a single \texttt{Other} class. The configuration sweep of Table~\ref{tab:sweep} suppresses only the \texttt{ip} cell and so never drops a row; to reproduce the EV row loss we apply equivalence-class-support suppression on the QI tuple, dropping every record whose class support is below $\tau$, the mechanism that took the EV release from 513 to 70 rows. We fix a stratified 70/30 split, suppress \emph{within} the training split, train a random forest ($n{=}400$, balanced class weights) on the survivors, and score the untouched test split (real arrivals are never suppressed). Table~\ref{tab:mlutility} sweeps $\tau$.

\begin{table}[t]
\centering
\caption{Downstream role-classification utility as equivalence-class suppression drives the surviving fraction down (664 pooled records; random forest, stratified 70/30, train on survivors, test on originals). $\tau$ is the class-support threshold; ``edge rec.''\ is macro-recall over the two rarest roles; ``F1 ret.''\ is macro-F1 relative to $\tau{=}1$.}
\label{tab:mlutility}
\renewcommand{\arraystretch}{1.0}
\setlength{\tabcolsep}{5pt}
\footnotesize
\begin{tabular}{@{}cccccc@{}}
\toprule
$\tau$ & survival & macro-F1 & accuracy & edge rec. & F1 ret.\\
\midrule
1  & $1.00$ & $0.69$ & $0.83$ & $0.92$ & $1.00$\\
2  & $0.52$ & $0.36$ & $0.74$ & $0.42$ & $0.53$\\
3  & $0.38$ & $0.35$ & $0.77$ & $0.42$ & $0.51$\\
5  & $0.24$ & $0.31$ & $0.75$ & $0.42$ & $0.45$\\
8  & $0.19$ & $0.26$ & $0.76$ & $0.42$ & $0.37$\\
12 & $0.14$ & $0.03$ & $0.12$ & $0.00$ & $0.04$\\
\bottomrule
\end{tabular}
\end{table}

Two things stand out. First, overall \emph{accuracy} is robust: it stays near $0.76$ even when suppression removes four fifths of the training rows ($\tau{=}8$, $19\,\%$ survival), because the dominant roles remain well represented among the survivors. An operator whose downstream model needs only the common function roles loses little. Second, this robustness is a majority-class effect. \emph{Macro}-F1, which weights every role equally, falls to $0.37$ of baseline, and recall on the two rarest roles halves from $0.92$ to $0.42$ once suppression begins. This is the edge-case degradation anticipated above: suppression preferentially deletes the sparse equivalence classes, where the minority roles live. Beyond $\sim$15\,\% survival ($\tau{=}12$, the EV regime) the surviving release is too skewed to train on and the model collapses. We therefore read suppression-heavy releases as adequate for majority-class analytics but unsafe for tasks that hinge on rare classes, and the dashboard surfaces the surviving fraction so an operator can see when a release has crossed into this regime. This degradation is avoidable in principle. As an exploratory, label-aware policy we exempt training records carrying the two rarest role labels from suppression: in the fixed split used here this restores edge-case recall from $0.42$ to $0.92$ and macro-F1 retention from $0.37$ to $0.68$ at $\tau{=}8$, while retaining more rows ($37\%$ vs $19\%$ survival). We evaluate it at $\tau{=}8$, the most severe regime before the classifier collapses at $\tau{=}12$. Because it consults the role label, it is an upper bound rather than a default release mechanism, and it trades a little privacy on rare records for their utility. A replication on the full structured EV set is future work.

\subsection{Discussion and Limitations}
\label{sub:limits}
The privacy layer is purely syntactic and does not yet offer a formal differential-privacy guarantee; it is subject to the known critiques of syntactic anonymisation~\cite{aggarwal2005,kifer2011}. It also treats each release in isolation: under \emph{sequential} re-publication of the same profiles (e.g.\ periodically released energy or EV-charging records), privacy can degrade through composition even when every individual release meets its target, a formal- and empirical-auditing problem we examine in complementary work~\cite{kekatos2026privacy}. The strongest policies are suppression-heavy: in the EV-charging case retention drops to $\sim 14\%$, and on the smallest PCAPs the surviving release reduces to one or two equivalence classes, a small-sample artefact rather than a meaningful release. The IPv4-shaped HMAC pseudonymisation is a controller-keyed heuristic, not a standards-grade FPE scheme such as AES-FF1. The O-RAN evaluation is a 12-session snapshot, and the 5GAD external corpus is benign-only; we have not yet evaluated PENM on attack-labelled, multi-UE, or partially-encrypted 5G/O-RAN traffic. The role classifier is rule-based and not quantitatively evaluated. The exact-QI measure reduces to $1/|E|$ by construction, so it is a full-QI syntactic bound rather than an independent attack; only the partial-QI and linkage results reflect simulated adversaries. The evaluation scale is modest throughout (tens of sessions in the O-RAN scenario, and tens to a few hundred records per capture), so our results are indicative of the pipeline's privacy/utility behaviour rather than a statistically powered study, and a larger multi-capture, multi-UE evaluation is scoped as future work. The downstream evaluation uses a single fixed stratified split; repeated resampling is future work. The utility-aware suppression result is label-aware and should be read as an exploratory upper bound, not a deployment-ready policy. The reported transformation latency covers only the privacy-quantification stage; profiler throughput is out of scope.

\section{Conclusions and Future Work}
\label{sec:conclusions}
This paper contributes an operator-side privacy-quantification layer for 5G/O-RAN network profiles: a syntactic-privacy baseline ($k$/$\ell$/$t$) aggregated into a composite Privacy-Posture Index, four pseudonymisation primitives, four generalisation profiles, and direct and threshold-based suppression, exposed through an inspectable dashboard. We evaluate it on an O-RAN testbed scenario ($k{=}3$, $\ell{=}2$, $t{=}0.25$) and on an EV-charging dataset, where the transformation raises $k$ from~1 to~5, $\ell$ from~1 to~2, and reduces $t$ from~0.90 to~0.73 at a $\sim$86\,\% record-suppression cost.

Several extensions remain. The privacy layer can be strengthened by partition-based $k$-anonymisation (Mondrian~\cite{lefevre2006}), third-generation syntactic models (Anatomy~\cite{xiao2006anatomy}, $m$-invariance~\cite{xiao2007minvariance}, $\delta$-presence~\cite{nergiz2007presence}), per-record disclosure-risk and centrality-based estimators~\cite{skinner2002,wagner2018,alemany}, and the formal $t$-closeness-to-DP and $k$-anonymity-to-DP bridges~\cite{domingoferrer2015tclose,li2012sampling} that would let PENM expose a uniform privacy budget. In practice we expect the harder extensions to be applied offline rather than inline~\cite{arshad}: differential privacy as an aggregate post-processing layer with R\'enyi-DP accounting~\cite{mironov2017}, federated learning with DP for cross-silo and hierarchical analytics, and zero-knowledge proofs with partially homomorphic encryption for cross-operator audit.

\section*{Acknowledgements}
This work has received funding from the European Union's Horizon Europe research and innovation programme under the Smart Networks and Services Joint Undertaking through the XTRUST-6G project (Grant Agreement No.~101192749). The authors thank the XTRUST-6G consortium partners for their feedback and collaboration, and the anonymous reviewers for their constructive comments and remarks.


\begin{thebibliography}{99}
\footnotesize

\bibitem{porambage2021} P.~Porambage \emph{et al.}, ``The roadmap to 6G security and privacy,'' \emph{IEEE Open J.~Commun.~Soc.}, vol.~2, pp.~1094--1122, 2021.
\bibitem{nguyen2022} V.-L.~Nguyen \emph{et al.}, ``Security and Privacy for 6G: A Survey on Prospective Technologies and Challenges,'' \emph{IEEE Commun.~Surveys Tuts.}, vol.~23, no.~4, pp.~2384--2428, 2021.
\bibitem{khan2020} R.~Khan \emph{et al.}, ``A survey on security and privacy of 5G technologies,'' \emph{IEEE Commun.~Surveys Tuts.}, vol.~22, no.~1, pp.~196--248, 2020.
\bibitem{polese2023} M.~Polese \emph{et al.}, ``Understanding O-RAN: Architecture, Interfaces, Algorithms, Security, and Research Challenges,'' \emph{IEEE Commun.~Surveys Tuts.}, vol.~25, no.~2, pp.~1376--1411, 2023.
\bibitem{enisa2019} ENISA, \emph{Pseudonymisation Techniques and Best Practices}, EU Agency for Cybersecurity, Tech.~Rep., Nov.~2019.
\bibitem{hintze} M.~Hintze and K.~El~Emam, ``Comparing the benefits of pseudonymisation and anonymisation under the GDPR,'' \emph{J.~Data Prot.~Priv.}, vol.~2, no.~2, pp.~145--158, 2018.
\bibitem{sweeney2002} L.~Sweeney, ``$k$-anonymity: A model for protecting privacy,'' \emph{Int.~J.~Uncertain.~Fuzziness Knowl.~Based Syst.}, vol.~10, no.~5, pp.~557--570, 2002.
\bibitem{samarati2001} P.~Samarati, ``Protecting respondents' identities in microdata release,'' \emph{IEEE Trans.~Knowl.~Data Eng.}, vol.~13, no.~6, pp.~1010--1027, 2001.
\bibitem{dalenius1977} T.~Dalenius, ``Towards a methodology for statistical disclosure control,'' \emph{Statistik Tidskrift}, vol.~15, pp.~429--444, 1977.
\bibitem{machanavajjhala2007} A.~Machanavajjhala, D.~Kifer, J.~Gehrke, and M.~Venkitasubramaniam, ``$\ell$-Diversity: Privacy beyond $k$-anonymity,'' \emph{ACM Trans.~Knowl.~Discov.~Data}, vol.~1, no.~1, p.~3, 2007.
\bibitem{li2007} N.~Li, T.~Li, and S.~Venkatasubramanian, ``$t$-Closeness: Privacy beyond $k$-anonymity and $\ell$-diversity,'' in \emph{Proc.~IEEE ICDE}, 2007, pp.~106--115.
\bibitem{lefevre2006} K.~LeFevre, D.~J.~DeWitt, and R.~Ramakrishnan, ``Mondrian multidimensional $k$-anonymity,'' in \emph{Proc.~IEEE ICDE}, 2006.
\bibitem{aggarwal2005} C.~C.~Aggarwal, ``On $k$-anonymity and the curse of dimensionality,'' in \emph{Proc.~VLDB}, 2005, pp.~901--909.
\bibitem{kifer2011} D.~Kifer and A.~Machanavajjhala, ``No free lunch in data privacy,'' in \emph{Proc.~SIGMOD}, 2011.
\bibitem{wagner2018} I.~Wagner and D.~Eckhoff, ``Technical privacy metrics: A systematic survey,'' \emph{ACM Comput.~Surv.}, vol.~51, no.~3, pp.~57:1--57:38, 2018.
\bibitem{arshad} R.~Arshad and M.~R.~Asghar, ``Characterisation and quantification of user privacy: Key challenges, regulations, and future directions,'' \emph{IEEE Commun.~Surveys Tuts.}, vol.~27, no.~5, pp.~3266--3307, 2025.
\bibitem{demontjoye2013} Y.-A.~de Montjoye, C.~A.~Hidalgo, M.~Verleysen, and V.~D.~Blondel, ``Unique in the crowd: The privacy bounds of human mobility,'' \emph{Sci.~Rep.}, vol.~3, p.~1376, 2013.
\bibitem{limniotis2024} K.~Limniotis, ``Cryptographic foundations of pseudonymisation for personal data protection,'' \emph{Cryptography}, vol.~10, no.~2, art.~18, 2026.
\bibitem{dwork2006} C.~Dwork, ``Differential privacy,'' in \emph{Proc.~ICALP}, 2006.
\bibitem{dwork2014} C.~Dwork and A.~Roth, ``The algorithmic foundations of differential privacy,'' \emph{Found.~Trends Theor.~Comput.~Sci.}, vol.~9, no.~3--4, pp.~211--407, 2014.
\bibitem{mironov2017} I.~Mironov, ``R\'enyi differential privacy,'' in \emph{Proc.~IEEE CSF}, 2017.
\bibitem{wei2020} K.~Wei \emph{et al.}, ``Federated learning with differential privacy,'' \emph{IEEE Trans.~Inf.~Forensics Secur.}, vol.~15, pp.~3454--3469, 2020.
\bibitem{aceto2019} G.~Aceto, D.~Ciuonzo, A.~Montieri, and A.~Pescap\'e, ``Mobile encrypted traffic classification using deep learning,'' \emph{IEEE Trans.~Netw.~Service Manage.}, vol.~16, no.~2, pp.~445--458, 2019.
\bibitem{maraj2026} D.~Maraj, R.~\v{S}oi\'{c}, A.~\v{Z}aja, and M.~Vukovi\'{c}, ``Privacy-preserving user profiling using MLP-based data generalization,'' \emph{Applied Sciences}, vol.~16, no.~2, art.~848, 2026.
\bibitem{shebaro2011} B.~Shebaro and J.~R.~Crandall, ``Privacy-preserving network flow recording,'' \emph{Digital Investigation}, vol.~8, suppl., pp.~S90--S100, 2011.
\bibitem{pang2003} R.~Pang and V.~Paxson, ``A high-level programming environment for packet trace anonymization,'' in \emph{Proc.~ACM SIGCOMM}, 2003, pp.~339--351.
\bibitem{anderson2017} B.~Anderson and D.~McGrew, ``Machine learning for encrypted malware traffic classification,'' in \emph{Proc.~ACM KDD}, 2017, pp.~1723--1732.
\bibitem{kamatchi2024} K.~Kamatchi and E.~Uma, ``Securing the edge: Privacy-preserving federated learning for insider threats in IoT networks,'' \emph{J.~Supercomput.}, vol.~81, no.~1, art.~246, 2025.
\bibitem{alemany} J.~Alemany, E.~del Val~Noguera, J.~Alberola~Oltra, and A.~Garc\'ia-Fornes, ``Estimation of privacy risk through centrality metrics,'' \emph{Future Gener.~Comput.~Syst.}, vol.~82, pp.~63--76, 2018.
\bibitem{idaholabs5gad} Idaho National Laboratory, ``5GAD: A 5G network traffic attack dataset,'' GitHub, 2022, accessed Jul.~2026. \url{https://github.com/IdahoLabResearch/5GAD}.
\bibitem{xiao2006anatomy} X.~Xiao and Y.~Tao, ``Anatomy: simple and effective privacy preservation,'' in \emph{Proc.~VLDB}, 2006, pp.~139--150.
\bibitem{xiao2007minvariance} X.~Xiao and Y.~Tao, ``M-invariance: Towards privacy preserving re-publication of dynamic datasets,'' in \emph{Proc.~ACM SIGMOD}, 2007, pp.~689--700.
\bibitem{nergiz2007presence} M.~E.~Nergiz, M.~Atzori, and C.~Clifton, ``Hiding the presence of individuals from shared databases,'' in \emph{Proc.~ACM SIGMOD}, 2007, pp.~665--676.
\bibitem{skinner2002} C.~J.~Skinner and M.~J.~Elliot, ``A measure of disclosure risk for microdata,'' \emph{J.~R.~Statist.~Soc.~B}, vol.~64, no.~4, pp.~855--867, 2002.
\bibitem{domingoferrer2015tclose} J.~Domingo-Ferrer and J.~Soria-Comas, ``From $t$-closeness to differential privacy and vice versa in data anonymization,'' \emph{Knowl.-Based Syst.}, vol.~74, pp.~151--158, 2015.
\bibitem{li2012sampling} N.~Li, W.~Qardaji, and D.~Su, ``On sampling, anonymization, and differential privacy: or, $k$-anonymization meets differential privacy,'' in \emph{Proc.~ACM ASIACCS}, 2012, pp.~32--33.
\bibitem{cvss40} FIRST, ``Common Vulnerability Scoring System version 4.0: Specification Document,'' Forum of Incident Response and Security Teams, 2023.
\bibitem{kekatos2026privacy} N.~Kekatos, M.~Korgiala-Karyda, and A.~Lekidis, ``Privacy specifications that do not compose: Empirical and formal auditing of sequentially published energy data,'' Research Square, preprint~rs-9984112/v1, 2026.

\end{thebibliography}
\end{document}